\documentclass[12pt,showpacs,prd]{revtex4}
\usepackage{epsfig,amssymb,amsfonts}

\makeatletter
\renewcommand{\@makefntext}[1]{\parindent=1em\noindent\hbox to 1.8em{\hss$^{\@thefnmark}$}#1}
\renewcommand{\@footnotemark}{\hbox{\mathsurround=0pt$^{\@thefnmark}$}}
\newcommand{\ftnote}[2]{\footnotemark[#1]\footnotetext[#1]{#2}}
\makeatother

\DeclareMathSymbol{\varGamma}{\mathord}{letters}{"00}

\begin{document}
\title{Chiral symmetry restoration in excited hadrons, quantum
fluctuations, and quasiclassics}
\author{L. Ya. Glozman}
\affiliation{Institute for Theoretical
Physics, University of Graz, Universit\"atsplatz 5, A-8010
Graz, Austria}
\author{A. V. Nefediev}
\affiliation{Institute of Theoretical and Experimental Physics, 117218,\\
B.Cheremushkinskaya 25, Moscow, Russia}
\author{J. E. F. T. Ribeiro}
\affiliation{ Centro de F\'\i sica das Interac\c c\~oes Fundamentais
(CFIF),Departamento de F\'\i sica, Instituto Superior T\'ecnico, Av.
Rovisco Pais, 1049-001 Lisbon, Portugal}
\newcommand{\be}{\begin{equation}}
\newcommand{\bea}{\begin{eqnarray}}
\newcommand{\ee}{\end{equation}}
\newcommand{\eea}{\end{eqnarray}}
\newcommand{\ds}{\displaystyle}
\newcommand{\low}[1]{\raisebox{-1mm}{$#1$}}
\newcommand{\loww}[1]{\raisebox{-1.5mm}{$#1$}}
\newcommand{\lmn}{\mathop{\sim}\limits_{n\gg 1}}
\newcommand{\vpint}{\int\makebox[0mm][r]{\bf --\hspace*{0.13cm}}}
\newcommand{\too}{\mathop{\to}\limits_{N_C\to\infty}}
\newcommand{\vp}{\varphi}
\newcommand{\vx}{{\vec x}}
\newcommand{\vy}{{\vec y}}
\newcommand{\vz}{{\vec z}}
\newcommand{\vk}{{\vec k}}
\newcommand{\vq}{{\vec q}}
\newcommand{\vpp}{{\vec p}}
\newcommand{\vn}{{\vec n}}
\newcommand{\vg}{{\vec \gamma}}

\begin{abstract}
In this paper, we discuss the transition to the semiclassical
regime in excited hadrons, and consequently, the restoration of
chiral symmetry for these states. We use a generalised
Nambu--Jona-Lasinio model with the interaction between quarks in
the form of the instantaneous Lorentz--vector confining potential.
This model is known to provide spontaneous breaking of chiral
symmetry in the vacuum via the standard selfenergy loops for
valence quarks. It has been shown recently that the effective
single--quark potential is of the Lorentz--scalar nature, for the
low--lying hadrons, while, for the high--lying states, it becomes
a pure Lorentz vector and hence the model exhibits the restoration
of chiral symmetry. We demonstrate explicitly the quantum nature
of chiral symmetry breaking, the absence of chiral symmetry
breaking in the classical limit as well as the transition to the
semiclassical regime for excited states, where the effect of
chiral symmetry breaking becomes only a small correction to the
classical contributions.
\end{abstract}
\pacs{12.38.Aw, 12.39.Ki, 12.39.Pn}
\maketitle

\section{Introduction}

Restoration of $SU(2)_L \times SU(2)_R$ and $U(1)_A$ chiral
symmetries of QCD in excited hadrons, both in baryons
\cite{G1,CG} and mesons \cite{G2,G3,G5}, was rather unexpected
(for a pedagogical overview see Ref.~\cite{G4}). Obviously, this
phenomenon requires a detailed experimental and theoretical study
and can provide a clue to our understanding of confinement, chiral
symmetry breaking, and their interrelation. Indeed, it has
recently become the subject of a significant theoretical effort
\cite{OPE,SWANSON,DEGRAND,SHIFMAN,parity2}.

As it is well known, in the chiral limit, both $SU(2)_L\times
SU(2)_R$ and $U(1)_A$ symmetries of the QCD Lagrangian are broken.
The spontaneous breaking of $SU(2)_L \times SU(2)_R \rightarrow
SU(2)_I$ \cite{NJL} is directly evidenced through a number of
facts, such as i) the absence of any multiplet structure of the
$SU(2)_L \times SU(2)_R$ group in the spectrum of low--lying
hadrons, ii) the Goldstone boson nature of the pion, iii) the
nonzero value of the quark condensate, which manifestly breaks
$SU(2)_L \times SU(2)_R$ symmetry of the vacuum, and through other
observables. The $U(1)_A$ symmetry breaking follows from the
absence of the multiplets of the $U(1)_A$ group low in the
spectrum and from a non-Goldstone nature of the $\eta'$-meson. The
$U(1)_A$ symmetry is broken by the axial anomaly of QCD
\cite{ANOMALY} and by the quark condensates in the vacuum
\cite{CG}. There is no one--to--one mapping of the low--lying
hadrons of positive and negative parity, thus indicating a strong
breaking of both $SU(2)_L \times SU(2)_R$ and $U(1)_A$ symmetries
and that chiral symmetry is realised nonlinearly in this part of
the hadronic spectrum.

However, already the first radial excitation of the pion,
$\pi(1300)$, has an approximately degenerate chiral partner,
$f_0(1370)$, which is predominantly a $\bar n n =
\frac{1}{\sqrt{2}}(\bar{u}u+\bar{d}d)$ state \cite{CZ}. The splitting within
the next radially excited doublet, $\pi(1800)$ --- $f_0(1790)$, is
already negligible\ftnote{1}{The $f_0(1790)$ state is clearly seen
as a $15\sigma$--peak $\bar n n$ state in the $J/\Psi$ decays
\cite{BES} and, previously, in $\bar p p$ collisions
\cite{BUGG}.}. There are further radial excitations of the $\pi$
and the $\bar n n$ $f_0$ states which are approximately
degenerate, as well as good chiral multiplets of higher--spin
mesons \cite{G2,G3}. Similarly, already in the $1.7GeV$ mass
region, the nucleon well established resonances demonstrate a
systematic pattern of an approximate parity doubling. An analogous
pattern persists also in the $\Delta$--spectrum, starting from the
$1.9GeV$ region, that is, at the same excitation energy with
respect to the ground state $\Delta(1232)$ as in the nucleon
spectrum. All these experimental facts indicate that the
restoration of chiral symmetry happens rather fast. Indeed, the
slowest possible rate of the symmetry restoration in mesons is
expected to be $\sim 1/n^{3/2}$, where $n$ is the radial quantum
number, as was recently shown in Ref.~\cite{SHIFMAN} through the
matching of the Operator Product Expansion and the resonance
representation of the two--point function.

There are quite unexpected implications of the chiral symmetry
restoration. For instance, the chiral partners of the excited
vector isovector $\rho$-mesons should be not only the
axial--vector isovector $a_1$ mesons (see, for example,
Ref.~\cite{OPE}), but also the axial--vector isoscalar $h_1$
mesons \cite{G3}. Consequently, there must be excited
$\rho$--mesons of two kinds, those which form $(1,0) \oplus (0,1)$
multiplets, together with the $a_1$--mesons, and those which fill
in $(1/2,1/2)$ chiral multiplets, together with the $h_1$--mesons
\cite{G3,SHIFMAN}. Actually, similar rules are valid also for all
possible mesons with $J \geqslant 1$ \cite{G3}. Consequently, all
highly excited hadrons can be viewed as colour--electric strings
with chiral quarks at the ends \cite{G5}.

While the role of different gluonic interactions of QCD, which
could be responsible for chiral symmetry breaking, is not yet
clear
--- these could be instantons, nonperturbative resummation of
perturbative gluon exchanges, gluodynamics responsible for the QCD
string formation, or something else, --- the most fundamental
reason of the chiral symmetry restoration in excited hadrons is
universal \cite{G6}.
 Namely, both $SU(2)_L \times SU(2)_R$ and $U(1)_A$ symmetries breaking results from quantum
fluctuations of the quark fields (that is, loops). However, for
highly excited states a quasiclassical regime necessarily takes
place\ftnote{2}{Some aspects of the semiclassical physics in
excited states, not related to the chiral symmetry restoration,
were discussed long ago --- see, for example, the bibliography in
Ref.~\cite{SHIFMAN}.}. Semiclassically, the contribution of
quantum fluctuations is suppressed, relative to the classical
contributions, by the factor of $\hbar/{\cal S}$, where ${\cal S}$
is the classical action of the intrinsic motion in the system
(that is, the full action of the hadron, in terms of quark and gluon
degrees of freedom, minus the action of the centre--of--mass
motion). Since, for highly excited hadrons, ${\cal S}\gg\hbar$,
then the symmetries of the classical Lagrangian must be restored.

While this argument is quite general and solid, it does not
provide us with any detailed microscopic picture of the symmetry
restoration. Then, in the absence of controllable analytic
solutions of QCD, such an insight can be obtained only through
models. It is the purpose of this paper to get such an insight.

Before discussing the details of the particular model to be
invoked, it is instructive to outline the minimal set of
requirements for such a model. It must be i) relativistic, ii)
chirally symmetric, and iii) able to provide a mechanism of
spontaneous breaking of chiral symmetry, iv) it should contain
confinement, in order to be able to address the issue of excited
states, and v) it must explain the restoration of chiral symmetry
for excited states. It is highly nontrivial indeed to meet all
these requirements with one and the same model. For example, the
famous Nambu and Jona-Lasinio model (NJL) \cite{NJL,REVIEWS} or
its specific realisations, like the instanton--liquid model
\cite{SHURYAK}, cannot be applied to excited hadrons as a matter
of principle, since they do not contain confinement as an
intrinsic feature, even though these models suggest an insight
into chiral symmetry breaking in the vacuum and into physics of
the lowest--lying hadrons. On the contrary, naive quark models,
with confinement postulated in the form of a raising potential
between the constituent quarks, do not contain any mechanism of
chiral symmetry breaking and hence must fail not only for the
low--lying hadronic states, but also for highly excited ones
\cite{G4,G5}.

There is a model, however, which does incorporate all required
elements. This is the generalised Nambu--Jona-Lasinio (GNJL) model
with the instantaneous vector confining kernel
\cite{Orsay,Orsay2,Lisbon}. This model is similar in spirit to the
large--$N_C$ 't~Hooft model in 1+1 dimensions \cite{tHooft} --- in
the latter confinement being provided by the linearly rising
Coulomb potential. An advantage of the GNJL model is that it can
be applied in 3+1 dimensions to systems of an arbitrary spin. In
this model, confinement of quarks is guaranteed due to the
instantaneous infinitely raising (for example, linear) potential,
which can be also supplied by extra ingredients, like the colour
Coulomb potential, for example. Then chiral symmetry breaking can
be described by the standard summation of the valence quark
selfinteraction loops (the mass--gap equation), while mesons are
obtained from the Bethe--Salpeter equation for the
quark--antiquark bound states. It was demonstrated in
Ref.~\cite{parity2} that, for the low--lying states, where chiral
symmetry breaking is important, this model leads to an {\it
effective} Lorentz--scalar binding interquark potential, while
for the high--lying states, such an effective potential becomes a
pure Lorentz spatial vector. As a result, the model does provide
chiral symmetry restoration. Besides, this model has an obvious
advantage as it is tractable and thus can be used as a laboratory
to get a microscopical insight into the restoration of chiral
symmetry for excited hadrons. It is the purpose of this paper to
demonstrate, in the framework of the GNJL model, the transition to
the quasiclassical regime for excited mesons and clarify the role
of quantum fluctuations in these systems.

\section{Generalised Nambu--Jona-Lasinio model}

\subsection{Introduction to the model and general remarks}

In this chapter, we give a short introduction to the GNJL chiral
quark model, called after the original paper \cite{NJL}, and which
is known to be a reliable testground for various low--energy
phenomena in QCD \cite{Orsay,Orsay2,Lisbon} . The model is
described by the Hamiltonian 
\be 
\hat{H}=\int d^3x\bar{\psi}(\vec{x},t)\left(-i\vec{\gamma}\cdot
\vec{\bigtriangledown}+m\right)\psi(\vec{x},t)+ \frac12\int d^3
xd^3y\;J^a_\mu(\vec{x},t)K^{ab}_{\mu\nu}(\vec{x}-\vec{y})J^b_\nu(\vec{y},t),
\label{H} 
\ee 
with the quark current--current
($J_{\mu}^a(\vec{x},t)=\bar{\psi}(\vec{x},t)\gamma_\mu\frac{\lambda^a}{2}
\psi(\vec{x},t)$) interaction parametrised by the instantaneous
confining kernel $K^{ab}_{\mu\nu}(\vec{x}-\vec{y})$ of a generic
form.

The model (\ref{H}), with the full kernel $K^{ab}_{\mu\nu}$
restricted to the $K^{ab}_{00}$ component only, was suggested in
the mid-eighties \cite{Orsay,Orsay2} and an instability of the
chirally--symmetric vacuum was demonstrated for the variety of
confining potentials of the form $|\vx|^\alpha$. Later, in the
series of papers \cite{Lisbon}, the given theory was rederived in
a generalised form through the introduction of Cooperlike $^3P_0$
quark-antiquark pairs, with the spatial components of the kernel,
$K^{ab}_{ij}$, included, which led to modifications of the
mass--gap equation. Constraints on the relative strengths of the
confining potentials of various Lorentz structure were imposed, in
order to maintain a divergence--free mass--gap equation. In
addition, the quark models of this class are demonstrated to
fulfill the well-known low--energy theorems, such as the
Gell-Mann--Oakes--Renner relation \cite{GOR} (for an early
derivation see, for example, Ref.~\cite{Orsay2}),
Goldberger--Treiman relation \cite{GT}, Adler selfconsistency
zero \cite{ASC}, the Weinberg theorem \cite{Wein} (derived for the
model (\ref{H}) in Ref.~\cite{EmilCota}), and so on. A convenient
analytic formalism to derive such low--energy theorems was
suggested in Ref.~\cite{BicudAp}, which is based on the approach
of Salpeter equations, put forward in the set of papers
\cite{Lisbon}. The axial anomaly and the $\pi\gamma\gamma$
coupling constant can be derived naturally in the framework of the
given model as well, following, for example, the lines of the
textbook \cite{CL} and using the asymptotic freedom and the Ward
identity for the dressed axial vertex \cite{BicudAp} (see also
Ref.~\cite{pi2g} for an independent derivation of this relation in
the Dyson--Schwinger approach to the dressed quark propagator).
Finally, the structure of the Hamiltonian (\ref{H}) can be traced
back to QCD in the Coulomb gauge \cite{coulg}.

In this paper, we use the simplest form of the kernel compatible
with the requirement of confinement \cite{Lisbon},
\be
K^{ab}_{\mu\nu}(\vec{x}-\vec{y})=g_{\mu 0}g_{\nu 0}\delta^{ab}V_0(|\vec{x}-\vec{y}|), 
\label{KK}
\ee
with the powerlike confining potential,
\be 
V_0(|\vec{x}|)=K_0^{\alpha+1}|\vec{x}|^{\alpha},\quad
0\leqslant\alpha\leqslant 2. 
\label{potential} 
\ee
Moreover, in order to make contact with the phenomenology of quarkonia, we
stick to the linearly growing potential (see, for example,
Ref.~\cite{linear}), which corresponds to the case $\alpha=1$ of
the general form (\ref{potential}). In the chiral limit, $m=0$,
the only remaining dimensional parameter is the strength of the
potential $K_0$ (in case of the linear confinement, we use the
conventional notation of the string tension $K_0^2=\sigma_0$). It
is remarkable, as mentioned in the introduction, that, in two
dimensions, the well--known 't~Hooft model for QCD$_2$
\cite{tHooft} in the large--$N_C$ limit and considered in the
axial gauge \cite{BG,2d}, also belongs to the theories of the
class (\ref{H}). In this case, the approximations of the
instantaneous nature of the interquark interaction and the
absence of higher--order kernels $K_{\mu\nu\lambda}^{abc},\ldots$
are exact and do not require any justification. In four
dimensions, such approximations are supported by independent
studies of the properties of the QCD vacuum. Indeed, the relative
smallness of the gluonic correlation length, measured on the
lattice \cite{Tg}, argues in favour of validity of the
instantaneous quark kernel, whereas the Casimir scaling, suggested
long ago \cite{cs1} and recently confirmed by lattice calculations
\cite{cs2}, supports the neglect of higher--order kernels
parametrising triple, quartic, and so on quark--current vertices
in the Hamiltonian (\ref{H}). Recently, the model (\ref{H}) was
considered from the point of view of the chiral symmetry
restoration high in the spectrum of mesons and, indeed, the parity
doubling was demonstrated to occur above the scale of about
$2.5GeV$, regardless of the explicit form of the confining
potential (\ref{potential}) \cite{parity2}. Although, in real QCD,
the chiral symmetry restoration in the hadronic spectrum seems to
occur somewhat faster --- possibly due to other nonconfining
contributions to the interquark interaction --- the found
estimate for the restoration scale is reasonable.

One can conclude, therefore, that the model under consideration,
as a beautiful testground for real QCD, provides a reliable source
of information about various phenomena related to chiral symmetry,
ranging from its spontaneous breaking in the lower part of
hadronic spectrum and up to its effective restoration high in the
spectrum. In this paper, we employ the model (\ref{H}) in order to
illustrate some general aspects of the phenomenon of spontaneous
breaking of chiral symmetry --- namely, to argue its intrinsically
quantum nature and to exemplify its restoration in the hadronic
spectrum as a general effect of restoration of the classical
symmetries of the fundamental QCD Lagrangian.

\subsection{Chiral symmetry breaking}

\begin{figure}[t]
\begin{center}
\epsfig{file=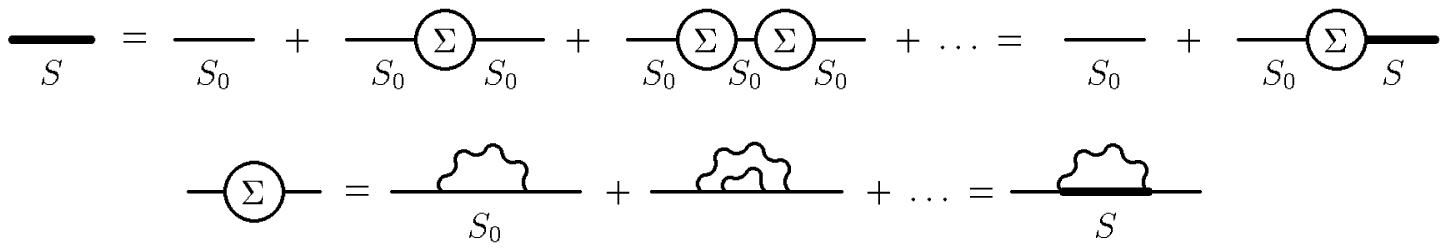,width=16cm} \caption{Graphical
representation of the equations for the dressed quark propagator,
Eq.~(\ref{De1}), and for the quark mass operator,
Eq.~(\ref{Sigma01}).}\label{diagrams}
\end{center}
\end{figure}

\begin{figure}[t]
\begin{center}
\epsfig{file=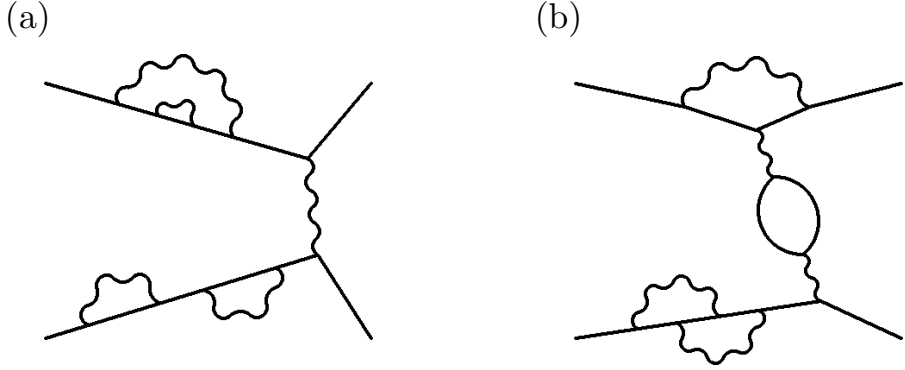,width=12cm} \caption{Planar (figure (a))
and nonplanar, suppressed by $N_C$, (figure (b))
diagrams.}\label{planar}
\end{center}
\end{figure}

The key idea of the BCS \cite{BCS} approach to spontaneous
breaking of chiral symmetry in the class of Hamiltonians described
by  Eq.~(\ref{H}) is the Dyson series for the quark propagator,
which takes the form, schematically: 
\be 
S=S_0+S_0\Sigma S_0+S_0\Sigma S_0\Sigma S_0+\ldots, 
\label{Ds} 
\ee 
with $S_0$ and $S$ being the bare-- and the dressed--quark propagators,
respectively, $\Sigma$ is the quark mass operator (see
Fig.~\ref{diagrams}). The series (\ref{Ds}) can be summed up to
produce the Dyson--Schwinger equation, 
\be 
S=S_0+S_0\Sigma S,
\label{De1} 
\ee 
with the formal solution being 
\be
S^{-1}(p_0,\vpp)=S_0^{-1}(p_0,\vpp)-\Sigma(\vpp), 
\label{Sm1} 
\ee
where the mass operator independence of the energy $p_0$ follows
from the instantaneous nature of the interaction --- see also
Eqs.~(\ref{Sigma01}) and (\ref{SiAB}) below. The expression for
the mass operator through the dressed--quark propagator (see
Fig.~\ref{diagrams}) reads: 
\be
i\Sigma(\vec{p})=C_F\int\frac{d^4k}{(2\pi)^4}V_0(\vec{p}-\vec{k})\gamma_0 S(\vec{k},k_0)\gamma_0,\quad 
C_F=\frac{N_C^2-1}{2N_C},
\label{Sigma01} 
\ee 
with both quark--quark--potential vertices
being bare momentum--independent vertices $\gamma_0$. This
corresponds to the so-called rainbow approximation which is well
justified in the limit of the large number of colours $N_C$. We
assume this limit in what follows. Then all nonplanar (such as
nonrainbow and those dressing the vertex $\gamma_0$) diagrams
appear suppressed by $N_C$ and can be consecutively removed from
the theory (see Fig.~\ref{planar}(a) for the examples of the
planar and Fig.~\ref{planar}(b) for nonplanar diagrams).
Eqs.~(\ref{De1}) and (\ref{Sigma01}) together produce a closed set
of equations, equivalent to a single nonlinear equation for the
mass operator, 
\be
i\Sigma(\vec{p})=\int\frac{d^4k}{(2\pi)^4}V(\vec{p}-\vec{k})\gamma_0\frac{1}{S_0^{-1}(k_0,\vk)-\Sigma(\vk)}\gamma_0, 
\label{Sigma03} 
\ee 
where the fundamental Casimir operator $C_F$ is absorbed to the potential,
$V(\vpp)=C_FV_0(\vpp)$, by introducing the fundamental string
tension $\sigma=C_F\sigma_0$.

As it often happens in the theories with strong interaction,
Eq.~(\ref{Sigma03}) may have multiple solutions. One of this
solutions is given by the perturbative series, schematically, 
\be
\Sigma=\int d^4k\;V\gamma_0 S_0\gamma_0+\int d^4k\;d^4q\;V^2\gamma_0 S_0\gamma_0 S_0 \gamma_0
S_0\gamma_0+\ldots, 
\label{Sser} 
\ee 
which converges fast in the weak interaction limit. Below we discuss in detail another,
nonperturbative, solution to Eq.~(\ref{Sigma03}). In the
meantime, alternatively, the series (\ref{Sser}) can be viewed as
an expansion in powers of the Planck constant $\hbar$. Indeed, the
mass operator is defined as the quark selfinteraction loop
integral which, being a purely quantum effect, is to be
proportional to $\hbar$. To see this explicitly, let us restore
$\hbar$'s in the expression for the mass operator (\ref{Sigma03}).
To this end, defining the confining potential through the area law
for the average of the closed Wilson loop (for completeness, we
restore here the speed of light $c$ as well), 
\be 
\langle W(C)\rangle=\exp\left[-\frac{\sigma A}{\hbar c}\right], 
\ee 
with $\sigma$ and $A$ being the string tension and the area of the
minimal Euclidean surface bounded by the contour $C$,
respectively, we extract the potential, for the rectangular loop
$C$: 
\be 
\frac{\sigma A}{\hbar c}=\frac{\sigma R\times (cT)}{\hbar c}=\frac{1}{\hbar}\int_0^T dt \sigma R=
\frac{1}{\hbar}\int_0^T dt\times {\rm potential}. 
\label{Wc} 
\ee
In the last formula, 
we restored both $\hbar$ and $c$, for the sake of transparency. However, in most formulae below, in order to simplify notations,
we do now show the speed of light $c$, restoring it only when necessary. Therefore, from Eq.~(\ref{Wc}), we
find the potential to be $V(r)=\sigma r$, which does not contain
$\hbar$ and is expected to remain finite in the formal classical
limit of $\hbar\to 0$. This constitutes the conjecture of
classical confinement, which we use throughout the paper. Now, the
Fourier transform of this potential is 
\be 
V(\vpp)=\int d^3 x e^{i\frac{\vpp\vx}{\hbar}} \sigma
|\vx|=-\frac{8\pi\sigma\hbar^4}{p^4}=\hbar^4 \tilde{V}(\vpp),
\label{FV} 
\ee 
where $\tilde{V}(\vpp)$ does not contain $\hbar$.
Then, in Eq.~(\ref{Sigma03}), the factor $\hbar^4$, from the
Fourier transform of the potential (\ref{FV}), will cancel
$\hbar$'s in the denominator, coming from
$\frac{d^4k}{(2\pi\hbar)^4}$, the only remaining Planck constant
being the one multiplying the entire integral in
Eq.~(\ref{Sigma03}), which is easily restored for dimensional
reasons. Thus, we find:
\be
i\Sigma(\vec{p})=\hbar\int\frac{d^4k}{(2\pi)^4}\tilde{V}(\vec{p}-\vec{k})
\gamma_0\frac{1}{S_0^{-1}(k_0,\vk)-\Sigma(\vk)} \gamma_0.
\label{Sigma02}
\ee
In other words, as easily seen from Eq.~(\ref{Sigma02}), each
power of the potential brings an extra loop integral and, as a
result, an extra power of $\hbar$.

To proceed, we use a convenient parametrisation of the mass
operator $\Sigma(\vpp)$, in the form \cite{Lisbon}: 
\be
\Sigma(\vec{p})=[A_p-m]+(\vec{\gamma}\hat{\vec{p}})[B_p-p],
\label{SiAB} 
\ee 
so that the dressed--quark Green's function (\ref{Sm1}) becomes 
\be
S^{-1}(\vec{p},p_0)=\gamma_0p_0-(\vec{\gamma}\hat{\vec{p}})B_p-A_p,
\label{SAB} 
\ee 
where, due to the instantaneous nature of the
interquark interaction, the time component of the four--vector
$p_\mu$ is not dressed.

It is easily seen from Eq.~(\ref{SAB}) that the functions $A_p$
and $B_p$ represent the scalar part and the space--vectorial part
of the effective Dirac operator. In the chiral limit, $A_p$
vanishes, unless chiral symmetry is broken spontaneously. It is
convenient, therefore, to introduce an angle, known as the chiral
angle $\vp_p$, according to the definition: 
\be
\tan\vp_p=\frac{A_p}{B_p}, 
\label{cha} 
\ee 
and varying in the
range $-\frac{\pi}{2}<\vp_p\leqslant\frac{\pi}{2}$, with the
boundary conditions $\vp(0)=\frac{\pi}{2}$, $\vp(p\to\infty)\to 0$.

The selfconsistency condition for the parametrisation
(\ref{SiAB}) of the nonlinear Eq.~(\ref{Sigma02}) requires that
the chiral angle obeys a nonlinear equation --- the mass--gap
equation, \be A_p\cos\vp_p-B_p\sin\vp_p=0, \label{mge2} \ee where
\be
A_p=m+\frac{\hbar}{2}\int\frac{d^3k}{(2\pi)^3}\tilde{V}(\vec{p}-\vec{k})\sin\vp_k,\quad
B_p=p+\frac{\hbar}{2}\int \frac{d^3k}{(2\pi)^3}\;(\hat{\vec{p}}\hat{\vec{k}})\tilde{V}(\vec{p}-\vec{k})\cos\vp_k.
\label{AB} 
\ee 
Notice that both functions $A_p$ and $B_p$ contain
classical and quantum contributions, the latter coming from loops.
For free particles, only the classical part survives and the
chiral angle (\ref{cha}) reduces to the free Foldy angle,
$\vp_p^{(0)}=\arctan\frac{m}{p}$, which diagonalises the free
Dirac Hamiltonian $H=\vec{\alpha}\vpp+\beta m$. The deep
connection between the chiral and the Foldy angles holds also for
nontrivial dynamically generated solutions to the mass--gap
Eq.~(\ref{mge2}) (see, for example, \cite{2d,NR1}).

The dispersive law of the dressed quark can be built then as 
\be
E_p=A_p\sin\vp_p+B_p\cos\vp_p, 
\label{Ep} 
\ee 
and it differs drastically from the free--quark energy, even becoming negative in
the low--momentum region. The latter behaviour is important in
order to explain a very small pion mass \cite{Orsay,Lisbon,NR1}.

\begin{figure}[t]
\begin{center}
\epsfig{file=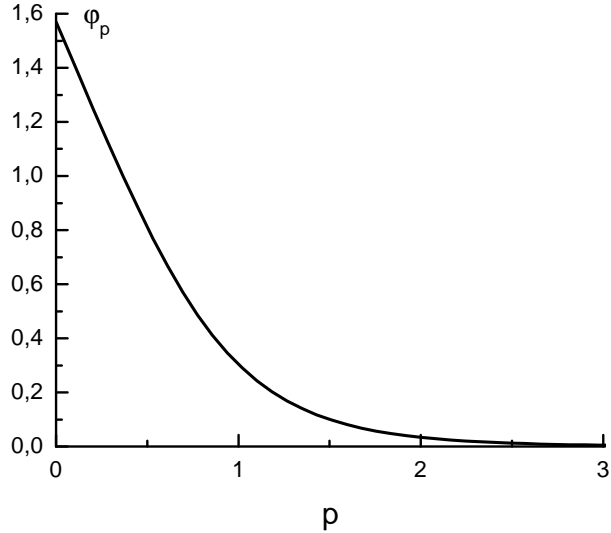,width=8cm} \caption{Nontrivial solution to the
mass--gap Eq.~(\ref{mge2}) with $m=0$ and for the linear
confinement (see, for example, Refs.~\cite{replica1,replica4} for the details). The
momentum $p$ is given in the units of $\sqrt{\sigma}$.}\label{vp}
\end{center}
\end{figure}

It was demonstrated in the pioneering papers \cite{Orsay} that,
for confining potentials, the mass--gap equation (\ref{mge2})
always possesses nontrivial solutions which break chiral symmetry
by generating a nontrivial masslike function $A_p$, even for
vanishing quark current mass. These solutions are given by smooth
decreasing functions which start at $\frac{\pi}{2}$ at the origin,
with the slope inversely proportional to the scale of chiral
symmetry breaking, generated by these solutions. At large momenta
they approach zero fast (see Fig.~\ref{vp}).

If the density of the vacuum energy is calculated for these
nontrivial solutions as
\be E_{\rm vac}[\vp_p]=-\frac{g}{2}\int\frac{d^3p}{(2\pi\hbar)^3}\left([A_p+m]
\sin\vp_p+[B_p+p]\cos\vp_p\right),\quad g=(2s+1)N_CN_f,
\ee
where the degeneracy factor $g$ counts the number of independent
quark degrees of freedom ($s=\frac12$ is the quark spin, $N_C$ and
$N_f$ are the number of colours and the number of flavours,
respectively), and is compared to the trivial, unbroken, vacuum
energy density, then one can see that, possessing a lower energy,
\be 
\Delta E_{\rm vac}=E_{\rm vac}[\vp_p]-E_{\rm vac}[\vp_p^{(0)}]<0, 
\ee 
the broken phase is energetically favourable and the unbroken phase is, therefore, unstable
\cite{Orsay}. The new, broken, vacuum is known as the BCS vacuum
of the theory. A detailed qualitative and  numerical investigation
of chiral symmetry breaking for various powerlike confining
potentials was performed in Ref.~\cite{replica4}.

\subsection{Breakdown of the  expansion for $\vp_p$ in powers of $\hbar$}

It was demonstrated, in the previous chapter, that all nontrivial
solutions to the mass--gap Eq.~(\ref{mge2}) appear entirely due to
loops and they are, therefore, of an intrinsically quantum nature.
In this chapter, we consider the mass--gap equation more closely
and study the expansion of the chiral angle in powers of $\hbar$.

For the sake of transparency, we write the mass--gap equation in
the explicit form and restore both the Planck constant $\hbar$ and
the speed of light $c$ in it,
\be
pc\sin\vp_p-mc^2\cos\vp_p=\frac{\hbar}{2}\int\frac{d^3k}{(2\pi)^3}
\tilde{V}(\vpp-\vk)\left[\cos\vp_k\sin\vp_p-
(\hat{\vec{p}}\hat{\vec{k}})\sin\vp_k\cos\vp_p\right]. 
\label{mgh}
\ee
Consider the chiral limit of $m=0$ first. Then, in the formal
classical limit of $\hbar\to 0$, the selfenergy (\ref{Sigma02})
as well as the right--hand side of Eq.~(\ref{mgh}) vanishes, and
the only solution of this equation is trivial, $\vp_p=0$. In view
of the discussion of the previous chapter, this result is obvious
since, being a genuine quantum entity --- parametrising loop
integrals --- $\vp_p$ should vanish in the classical limit. A
naive attempt to search for solutions of Eq.~(\ref{mgh}) in the
form of the formal expansion in $\hbar$, $\vp_p=\hbar\times
f_1(p)+\hbar^2\times f_2(p)+\ldots$, fails --- all coefficients
vanish. The reason of such a failure can be considered from two
viewpoints. Formally, to perform such an expansion, one is to
build a quantity, $\cal S$, with the dimension of the classical
action, in order to perform the actual  expansion,
$\vp_p=\frac{\hbar}{\cal S}\times f_1(p)+\frac{\hbar^2}{{\cal
S}^2}\times f_2(p)+\ldots$. With only two dimensional parameters
in hand, the string tension $\sigma$ and the speed of light $c$,
it is not possible to build such an action. On the other hand, one
can arrive at the same conclusion by a direct inspection of
Eq.~(\ref{mgh}). Indeed, the chiral angle is a dimensionless
function of the form $\vp_p=f_0(p/(\mu c))$, where the mass scale
$\mu$ is to be built out of the dimensional parameters of the
theory. In order to build this scale, let us introduce
dimensionless variables in the integral, $\vpp=\mu c\vx$ and
$\vk=\mu c\vy$, and define $\mu$ such that the resulting equation
defines only the profile of the chiral angle and does not contain
any scale at all. The result is $\mu=\sqrt{\sigma\hbar c}/c^2$ and we
end up with a nonanalytic dependence of $\vp_p$ on the Planck
constant. This is easily seen, for example, from the low--momentum
expansion of $\vp_p$,
$$
\vp_p\mathop{\approx}\limits_{p\to 0}\frac{\pi}{2}-{\rm
const}\frac{p c}{\mu c^2}+ \ldots=\frac{\pi}{2}- {\rm
const}\frac{pc}{\sqrt{\sigma\hbar c}}+\ldots.
$$

As $\hbar$ vanishes, the chiral angle gets steeper and steeper at
the origin, approaching the trivial solution $\vp_p=0$ for all
$p$'s and, in the limit of $\hbar\rightarrow 0$, we cease to have
a low--momentum expansion of $\vp_p$. Such a collapse of the
chiral angle in the classical limit has a rather deep physical
reason. Indeed, the pseudounitary operator creating the broken
BCS vacuum $|0\rangle$ from the unbroken vacuum $|0\rangle_0$ is
\cite{Lisbon,replica2} 
\be 
|0\rangle=U_0|0\rangle_0,\quad
U_0=e^{Q^\dagger-Q},\quad
Q^\dagger=\frac12\int\frac{d^3p}{(2\pi\hbar)^3}\vp_pC_p^\dagger,
\label{S0} 
\ee 
\be
C_p^\dagger=[b^\dagger_{\uparrow}(\vec{p}),b^\dagger_{\downarrow}(\vec{p})]
\mathfrak{M}_{^3P_0}
\left[\begin{array}{c}d^\dagger_{\uparrow}(\vec{p})\\
d^\dagger_{\downarrow}(\vec{p})\end{array}\right],\quad
\mathfrak{M}_{^3P_0}=(\vec{\sigma}\hat{\vec{p}})i\sigma_2,
\label{Cddef} 
\ee 
where $\sigma$'s are the $2\times 2$ Pauli
matrices. Therefore, the chiral angle is nothing but the wave
function of the BCS $^3P_0$ quark--antiquark pairs in the vacuum,
created with the operator $C_p^\dagger$. Besides that, the chiral
angle (in the form of $\sin\vp_p$) defines the wave function of
the chiral pion \cite{Orsay,Lisbon,NR1}. In the classical limit,
wave functions of  quantum systems are known to collapse and no
expansion in powers of $\hbar$ is possible\ftnote{3}{Consider, for
example, the one--dimensional nonrelativistic harmonic oscillator
wave functions, which behave as $\psi_n(x)\propto
\exp\left[-\frac{m_0\omega x^2}{\hbar}\right]$, $m_0$ and $\omega$
being the oscillator mass and frequency, respectively. Although
the pre-exponential factor in $\psi_n$ also contains $\hbar$, the
exponent dominates in the limit $\hbar\to 0$ destroying the wave
functions for all finite $x$'s.}. We arrive, therefore, at the
purely \lq\lq nonperturbative" in $\hbar$ solution to the
mass--gap equation and thus an inevitable breakdown of the
expansion  of the chiral angle $\vp_p$ in $\hbar$. It is
instructive to go now slightly beyond the chiral limit and to
switch on a small current quark mass $m$. Working on top of the
chirally nonsymmetric BCS vacuum, one can incorporate a small $m$
as a perturbation, in the spirit of the standard chiral
perturbation theory, building an infinite series of corrections to
the leading regime $\vp_p=f_0(pc/\sqrt{\sigma\hbar c})$, suppressed
by the small parameter $\frac{mc^2}{\sqrt{\sigma\hbar c}}$,
\be
\vp_p=\sum_{n=0}^\infty\left(\frac{mc^2}{\sqrt{\sigma\hbar c}}\right)^nf_n
\left(\frac{pc}{\sqrt{\sigma\hbar c}}\right), 
\label{vp1} 
\ee
where the leading term --- the function $f_0$ --- is known only
numerically and is depicted in Fig.~\ref{vp}.

However, beyond the chiral limit, the nonanalytic in $\hbar$
behaviour of the chiral angle is \lq\lq smeared" by the quark
mass, so that $\vp_p$ does support now an  expansion in $\hbar$
--- this can be verified by a direct investigation of the
mass--gap Eq.~(\ref{mgh}). One can also arrive at the same
conclusion using general qualitative arguments. Indeed, for a
nonvanishing quark mass $m$, it is possible to build the quantity
with the dimension of the classical action, 
${\cal S}\sim\frac{m^2c^3}{\sigma}$, so that the actual parameter of the
expansion appears to be $\frac{\sigma\hbar c}{(mc^2)^2}$ and the
mass--gap Eq.~(\ref{mgh}) admits a solution in the form of a
\lq\lq perturbative" series in powers of $\hbar$, 
\be
\vp_p=\sum_{n=0}^\infty\left(\frac{\sigma\hbar c}{(mc^2)^2}\right)^n\tilde{f}_n\left(\frac{p}{mc}\right),
\label{vp2} 
\ee 
the leading term giving just the free solution,
$\tilde{f}_0 = \arctan\frac{mc}{p}$. As a matter of fact, this
\lq\lq perturbative" solution is represented by the series for the
quark selfenergy (\ref{Sser}).

It is easy to notice that the two expansions given by
Eqs.~(\ref{vp1}) and (\ref{vp2}) follow from one and the same
solution to the mass--gap equation. However, while the expansion
(\ref{vp1}) is appropriate around the chiral limit $m=0$ and
beyond the classical limit, $\hbar\neq 0$ (the expansion
(\ref{vp2}) fails in this region), for
$m\gg\frac{\sqrt{\sigma\hbar c}}{c^2}$, the expansion (\ref{vp2}) is
much better than that of Eq.~(\ref{vp1}). Notice  that there is no
one--to--one correspondence between the sets of functions
$\{f_n\}$ and $\{\tilde{f}_n\}$ (for example, the function $f_0$,
depicted in Fig.~\ref{vp}, approaches zero as $1/p^5$ at large
$p$'s, whereas the asymptotic behaviour of $\tilde{f}_0$ is much
slower, only $1/p$).

In conclusion, notice that Eqs.~(\ref{vp1}) and (\ref{vp2}) explicitly define two 
different regimes, according to the value of the parameter
$\frac{m}{\sqrt{\sigma}}$. Chiral symmetry and its spontaneous breaking
are relevant for $m\ll\sqrt{\sigma}$ --- the regime (\ref{vp1}), while the 
\lq\lq heavy--quark physics" is adequate in the opposite limit of 
$m\gg\sqrt{\sigma}$ --- the regime (\ref{vp2}).

\section{Parity doubling for highly excited hadrons}

\begin{figure}[t]
\begin{center}
\epsfig{file=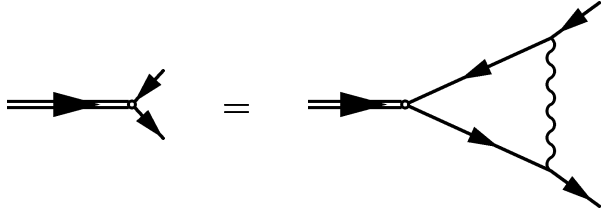,width=8cm} \caption{Graphical representation
of the Bethe--Salpeter Eq.~(\ref{GenericSal}) for the
quark--antiquark bound state, in the ladder
approximation.}\label{bseq}
\end{center}
\end{figure}

In this chapter, we consider the spectrum of heavy--light mesons
and demonstrate explicitly the quantum nature of the splitting
within chiral doublets. To this end, we derive the bound--state
equation for the quarkonium made of a static antiquark, placed at
the origin, and a light quark. We follow the lines of
Ref.~\cite{parity2}.

The Bethe--Salpeter equation for the mesonic Salpeter amplitude
$\chi({\vec p};M)$, depicted graphically in Fig.~\ref{bseq},
reads: 
\be 
\chi({\vec p};M)=-i\int\frac{d^4k}{(2\pi)^4}V(\vec{p}-\vec{k})\; \gamma_0
S_q({\vec k},k_0+M/2)\chi({\vec k};M)S_{\bar Q}({\vec k},k_0-M/2)\gamma_0, 
\label{GenericSal} 
\ee 
where the light--quark propagator (\ref{Sm1}) (we use the subscript $q$ in order to
distinguish it from the static--particle propagator, labelled by
$\bar{Q}$) can be written as a sum of the positive-- and
negative--energy parts \cite{Lisbon,kw}, 
\be 
S_q({\vec p},p_0)=\frac{\Lambda^{+}({\vec p})\gamma_0}{p_0-E_p+i\epsilon}+\frac{{\Lambda^{-}}({\vec
p})\gamma_0}{p_0+E_p-i\epsilon}, 
\label{Feynman} 
\ee 
with the dressed projectors being 
\be 
\Lambda^\pm(\vec{p})=T_pP_\pm T_p^\dagger,\quad P_\pm=\frac{1\pm\gamma_0}{2},\quad
T_p=\exp{\left[-\frac12(\vec{\gamma}\hat{\vec{p}})\left(\frac{\pi}{2}-\vp_p\right)\right]}.
\ee 
Eq.~(\ref{GenericSal}) is written in the ladder approximation
which, similarly to the rainbow approximation for the quark mass
operator, is well controlled in the large--$N_C$ limit.

Similarly, for the static particle, the chiral angle is simply $\vp_{\bar Q}(p)=\frac{\pi}{2}$,
so that its Green's function takes a very simple form,
\be
S_{\bar Q}({\vec p},p_0)=\frac{P_+\gamma_0}{p_0-m_{\bar Q}+i\epsilon}+
\frac{P_-\gamma_0}{p_0+m_{\bar Q}-i\epsilon}.
\ee

Now, defining the vertex $\varGamma(\vec{p})$, rotated with the
light--quark Foldy operator $T^\dagger_p$ \cite{kw} from the left
and with the heavy--quark Foldy operator, reduced to unity, from
the right,
\be
\varGamma(\vec{p})=T_p^\dagger\left[\int\frac{dp_0}{2\pi}S_q(\vec{p},p_0+M/2)\chi(\vec{p};M)S_{\bar
Q}(\vec{p},p_0-M/2)\right]\hat{1},
\ee
and performing the energy integration in Eq.~(\ref{GenericSal}), we arrive at the matrix
equation for the new vertex,
\be
(E-E_p)\varGamma(\vec{p})=P_+\left[\int\frac{d^3k}{(2\pi)^3}V(\vec{p}-\vec{k})T_p^\dagger
T_k\varGamma(\vec{k})\right]P_-. 
\label{psip}
\ee
For the sake of convenience, we defined the energy excess over the
static--particle mass, $E=M-m_{\bar Q}$. Due to the projectors on
the right--hand side of Eq.~(\ref{psip}), the matrix $\varGamma(\vec{p})$ should be searched in the form
\be
\varGamma(\vec{p})= \left(
\begin{array}{cc}
0&\psi(\vec{p})\\
0&0
\end{array}
\right)={\psi(\vec{p})\choose 0}_q\otimes(0\;1)_{\bar Q},
\label{qbQ} 
\ee
which is valid for $M>0$. For negative $M$'s, when the
quark and the antiquark are interchanged, the matrix
$\varGamma(\vec{p})$ obeys an equation similar to Eq.~(\ref{psip})
and takes the form:
\be
\varGamma(\vec{p})= \left(
\begin{array}{cc}
0&0\\
\psi(\vec{p})&0
\end{array}
\right)={0\choose\psi(\vec{p})}_{\bar q}\otimes(1\;0)_Q.
\label{bqQ} 
\ee
Thus, in view of a very passive role played by the static
constituent, the heavy--light meson wave function $\Psi(\vec{p})$
can be identified with the light--particle wave function \cite{2d,parity2},
$$
\Psi(\vec{p})_{|M>0}={\psi(\vec{p})\choose 0},\quad \Psi(\vec{p})_{|M<0}={0\choose\psi(\vec{p})},
$$
where $\psi(\vpp)$ obeys a Schr{\" o}dingerlike equation \cite{parity2},
\be
E_p\psi(\vec{p})+\int\frac{d^3k}{(2\pi)^3}V(\vpp-\vk)\left[C_pC_k+
({\vec \sigma}\hat{\vpp})({\vec
\sigma}\hat{\vk})S_pS_k\right]\psi(\vec{k})=E\psi(\vec{p}),
\label{FW4}
\ee
with $C_p=\cos\frac12\left(\frac{\pi}{2}-\vp_p\right)$ and
$S_p=\sin\frac12\left(\frac{\pi}{2}-\vp_p\right)$.

The bound--state Eq.~(\ref{FW4}) was studied in detail in
Ref.~\cite{parity2} and it was demonstrated to support parity
doublers high in the spectrum of heavy--light bound states.
Moreover, a relation between chiral symmetry restoration for
highly excited states and the Lorentz nature of the effective
interquark interaction was established in the same work.
Following the lines of the cited paper and performing the reversed
Foldy--Wouthuysen transformation, with the help of the operator
$T_p$, one can derive a Diraclike effective one--particle
equation for the light--quark Foldy--counter--rotated wave
function, $\tilde{\Psi}(\vec{p})=T_p\Psi(\vec{p})$, in coordinate
space, 
\be 
(c{\vec \alpha}\vpp+\beta mc^2)\tilde{\Psi}(\vx)+\frac{\sigma}{2}\int d^3z
\left(\vphantom{A^2}|\vx|+|\vz|-|\vx-\vz|\right)U(\vx-\vz)\tilde{\Psi}(\vz)=E\tilde{\Psi}(\vx),
\label{DS2} 
\ee 
with the unitary matrix 
\be
U(\vx)=\int \frac{d^3p}{(2\pi\hbar)^3}U(\vpp)e^{-i\frac{\vpp\vx}{\hbar}},\quad
U(\vpp)=\beta\sin\vp_p+({\vec\alpha}\hat{\vpp})\cos\vp_p.
\label{L2} 
\ee

We have restored, in Eqs.~(\ref{DS2}) and (\ref{L2}), the speed of
light and the Planck constant. The Lorentz nature of the effective
interquark interaction in the Diraclike Eq.~(\ref{DS2}) is
governed by the structure of the matrix $U$, that is, by the value
of the chiral angle $\vp_p$. Indeed, for
$\vp_p\approx\frac{\pi}{2}$, the effective interaction is scalar,
so that no parity doublers can appear. On the contrary, for a
vanishing chiral angle, the interaction becomes vectorial,
Eq.~(\ref{DS2}) respects chiral symmetry and, as a result, this
symmetry manifests itself in the spectrum, in the form of
degeneracy in mass between the states with opposite parity. This
obviously happens to highly excited mesons, since the mean
relative interquark momentum in such states is large, and,
consequently, the corresponding value of the chiral angle is small
(see Fig.~\ref{vp}). Notice, however, that as discussed above, in
the chiral limit, a nontrivial angle $\vp_p$ appears entirely due
to quantum fluctuations (loops) and it is a function of the
argument $pc/\sqrt{\sigma\hbar c}$. Then, considering large
relative momenta is equivalent to taking the formal classical
limit of $\hbar\to 0$. In this limit, it follows from the
mass--gap equation that $\sin\vp_p=0$, consequently
Eq.~(\ref{DS2}) becomes completely classical with the
Lorentz--vector spatial potential. We see, therefore, that, in
agreement with general expectations \cite{G6}, the effects of
quantum fluctuations fade out in the semiclassical region of
highly excited hadrons --- chiral symmetry restoration in the
spectrum being a clear manifestation of this fact. Remarkably, the
GNJL model (\ref{H}) provides quite natural surroundings for the
investigation of this phenomenon.

\section{Conclusions}

In this paper, in the framework of the generalised NJL model with
the Lorentz--vector instantaneous interquark interaction, we
illustrate explicitly the effect of quantum fluctuations
suppression for highly excited states in the hadronic spectrum
and, consequently, the chiral symmetry restoration in these
states. The given model is known to be a source of information on
various phenomena related to chiral symmetry in QCD, ranging from
its spontaneous breaking in the BCS vacuum, through the standard
selfenergy loops for valence quarks, to its effective restoration
for highly--excited hadrons. It was demonstrated recently that the
effective single--quark potential in the heavy--light quarkonium
changes its Lorentz nature from a pure scalar, for the low--lying
bound states, to a pure spatial vector, for highly excited
hadrons, thus providing a clear pattern of the chiral symmetry
restoration in the spectrum.

We closely study the mass--gap equation for this theory in the
chiral limit and demonstrate explicitly spontaneous breaking of chiral
symmetry to be a purely quantum effect, originating from the
nonperturbative summation of the selfenergy loops for quarks.
The nontrivial solution to the mass--gap equation vanishes in the
classical limit and the chiral angle possesses a peculiar,
nonanalytic, dependence on the Planck constant,
$\vp_p=f(pc/\sqrt{\sigma\hbar c})$ (plus corrections in powers of
the expansion parameter $\frac{mc^2}{\sqrt{\sigma\hbar c}}$, if
a small current quark mass is introduced on top of the chirally
nonsymmetric BCS vacuum). As a result, no  expansion of the chiral
angle in powers of $\hbar$ is possible in the chiral limit.

Furthermore, the form of the bound--state equation for the
quark--antiquark meson --- for the sake of simplicity we consider
a heavy--light state --- suggests that, in the chiral limit, the
Lorentz--scalar part of the effective interquark interaction is
proportional to $\sin\vp_p$ and, therefore, it is of a purely
quantum origin (it originates from selfenergy loops).
Consequently, for highly excited hadrons, possessing a large
relative momentum, this scalar part of the interaction vanishes,
and only the classical Lorentz--vector part survives, giving rise
to the degeneracy in mass of states with the opposite parity. This
conclusion provides an explicit realisation of the general
statement that all quantum loop effects must disappear for highly
excited hadrons, where the semiclassical regime necessarily takes
place, with dominating contributions being purely classical and
all quantum loop effects giving only small corrections. This
provides the most general and solid argument in favour of the
chiral symmetry restoration in the upper part of the hadronic
spectrum.

\begin{acknowledgments}
L. Ya. G. thanks M. Shifman for comments and acknowledges the support from the P16823-N08 project
of the Austrian Science Fund. A. V. N. would like to acknowledge useful discussions with Yu. S. Kalashnikova and the
financial support of the grants DFG 436 RUS 113/820/0-1, RFFI 05-02-04012-NNIOaÁ, 
and NS-1774.2003.2, as well as of the
Federal Programme of the Russian Ministry of Industry, Science, and
Technology No 40.052.1.1.1112.
\end{acknowledgments}

\end{document}